\documentstyle[11pt,twoside,colloqOHP,epsfig]{article}
\begin{document}
\title{Abundance ratios of volatile vs.\ refractory elements: hints of pollutions?}
\author{Ecuvillon$^1$, G. Israelian$^1$, N. C. Santos$^{2,3}$, M. Mayor$^3$, G. Gilli$^{1}$}
\affil{$^1$Instituto de Astrofisica de Canarias, La Laguna, Tenerife, Spain}
\affil{$^2$Observatorio Astronomico de Lisboa, Lisboa, Portugal}
\affil{$^3$Observatoire de Gen\'eve, 1290 Sauverny, Switzerland}
\begin{abstract}
We present the abundance ratios [X/H] of a large set of chemical species with T$_C$ from 75 to 1600\,K in an almost complete set of 105 planet-host stars and in a volume-limited comparison sample of 88 stars without any known planets. The large range of different T$_C$ covered by all the analysed elements allows us to investigate possible anomalous trends of [X/H] vs.\ T$_C$ in targets with planets with respect to comparison sample stars. This can give important hints for the detection of pollution events and for the understanding of the relative contribution of the differential accretion to the average metallicity excess found in planet host stars.
\end{abstract}
\section{Introduction}
Two main explanations have been suggested to link the average iron excess found in planet host stars (Santos et al.\ 2005, and references therein) to the presence of planets: the self-enrichement scenario (Gonzalez 1997) and the primordial hypothesis (Santos, Israelian \& Mayor 2000, 2001). If the star underwent significant pollution, refractory elements might be added preferentially compared to volatiles, due to the high temperature environment, and thus abundance patterns depending on the elemental condensation temperature T$_C$ should be observed. Smith, Cunha, \& Lazzaro (2001) found a subset of 5 planet-host stars which exhibited an accentuated trend of increasing [X/H] with T$_C$. Nevertheless, their analysis used an inhomogeneous set of abundances from different authors. Subsequent studies by Takeda et al.\ (2001) and Sadakane et al.\ (2002) obtained no peculiar trends of [X/H] with T$_C$ in a sample of 14 and 12 stars with planets, respectively. 
\begin{figure}[h]
\center
\epsfig{file=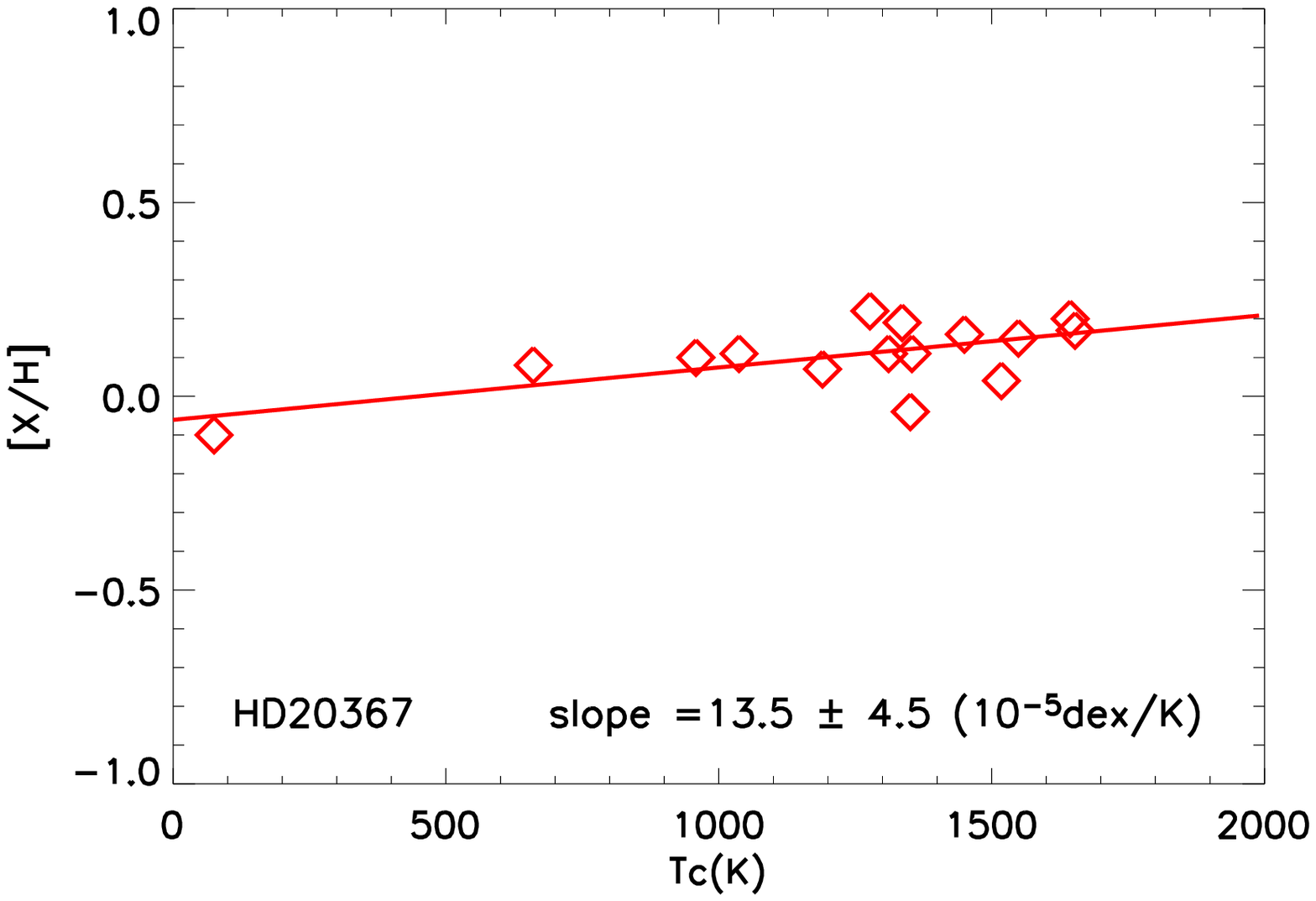,width=5cm}
\epsfig{file=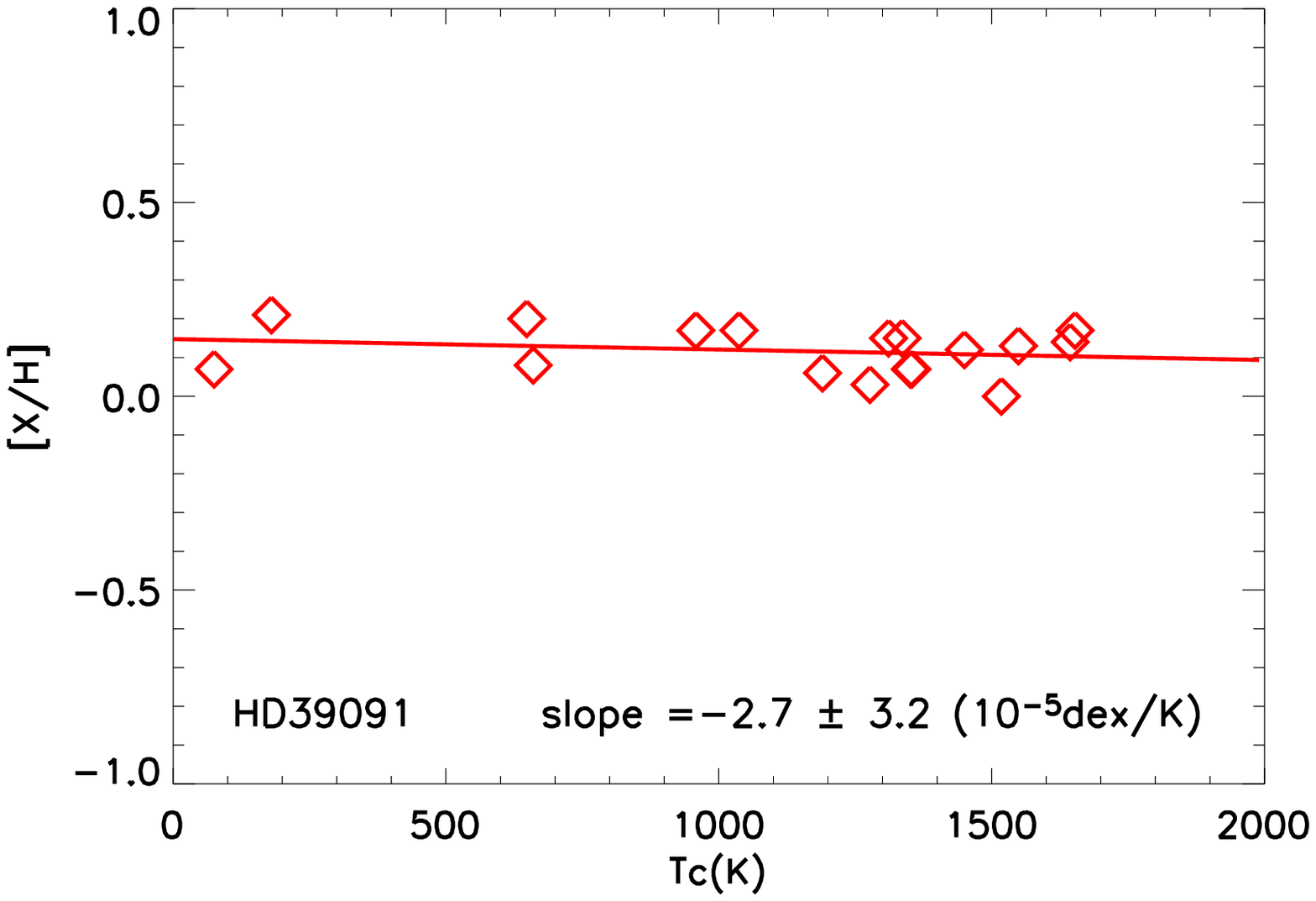,width=5cm}
\caption{[X/H] vs. T$_C$ for the planet host stars HD\,20387 and HD\,39091. The solid lines represent the linear least-square fits to the data and the slope values are written at the bottom of each plot.}
\end{figure}
\section{Analysis and Results}
We derived abundance ratios in an almost complete set of 105 planet-host stars and in a volume-limited comparison sample of 88 stars, for the volatile elements CNO, S and Zn (see Ecuvillon et al.\ 2004a, 2004b, 2005) and for the refractories: Cu, Si, Ca, Sc, Ti, V, Cr, Mn, Co, Ni, Na, Mg, Al (see Ecuvillon et al.\ 2004b; Beirao et al.\ 2005; Gilli et al.\ 2005).
For each target the [X/H] vs.\ T$_C$ trend and the corresponding linear least-square fit were obtained (see Fig.\,1). We excluded from our analysis targets with less than 14 abundance determinations and less than 2 volatiles, in order to span a large range of T$_C$ in each case. In Fig.\,2 (left panel), a subsample of 18 stars, 16 and 2 with and without planets, respectively, shows slopes larger than the average plus sigma.
\begin{figure}[h] 
\epsfig{file=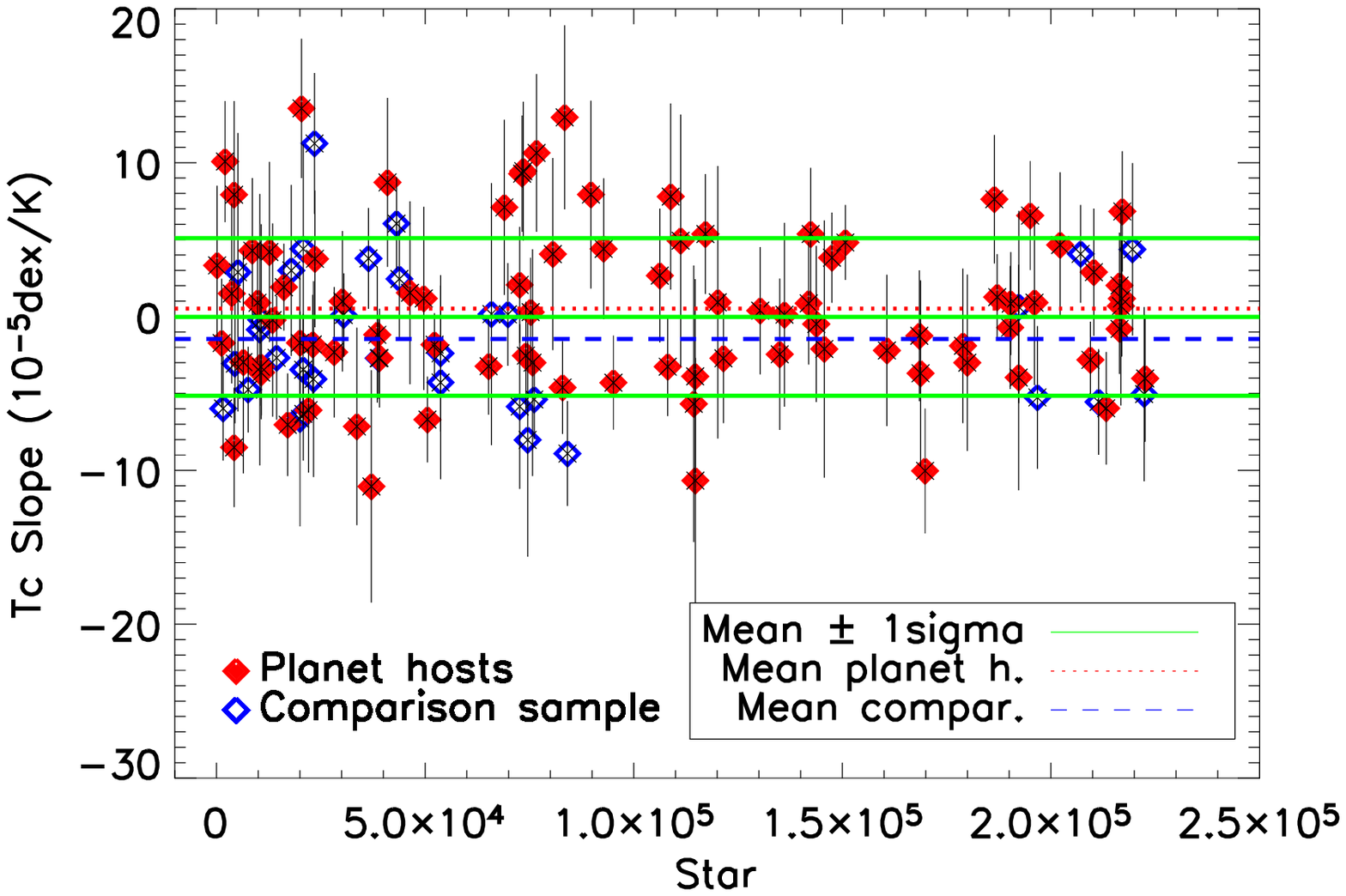,width=5.9cm}
\epsfig{file=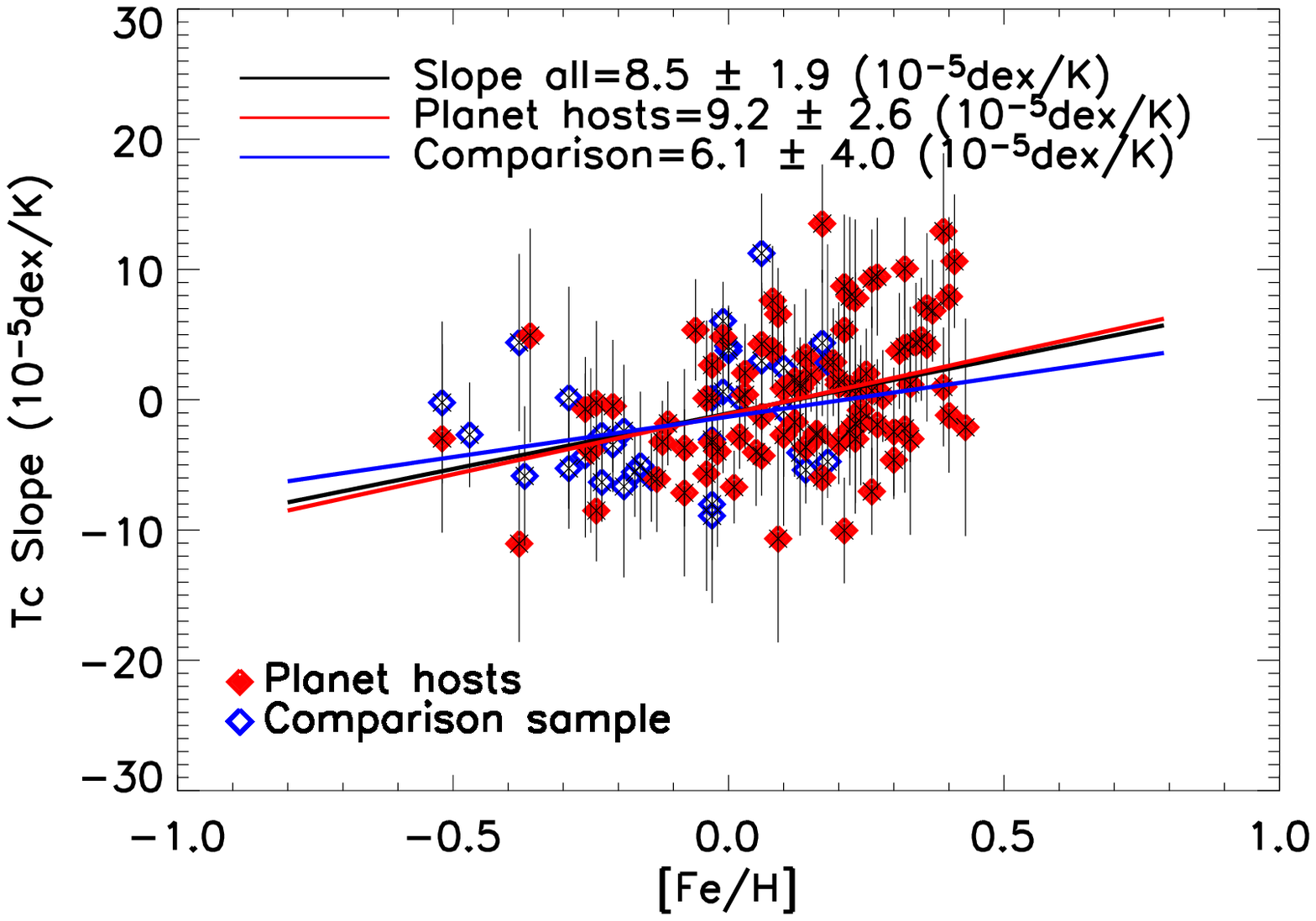,width=5.9cm}
\caption{Slopes derived from [X/H] vs. T$_C$ for planet host (red filled symbols) and comparison sample stars (blue open symbols). The green solid, red dotted and blue dashed lines represent the mean slopes (upper panel) and the linear least-square fits (lower panel) for all the targets of the two samples, and for the stars with and without planets, respectively.}
\end{figure}

\section{Discussion and Conclusions}
Our results do not reveal any striking difference in the [X/H] vs.\,T$_C$ trends between stars with and without planets. A subsample of 16 stars with and 2 without planets, with slopes larger than the average value plus sigma, might bear the signature of having undergone some pollution event (see Fig.\,2, left panel). In fact, the accretion of planetesimal material in a high temperature environment might produce larger [X/H] vs.\,T$_C$ slopes. All the stars display an increasing trend of the T$_C$ slopes with increasing [Fe/H] (see Fig.\,2, right panel), due to the Galactic chemical evolution. However, planet-host stars with [Fe/H]$>$0.1 present a steeper linear trend with respect to the comparison sample. This would suggest that pollution events are much likelier to occur at super-solar [Fe/H]. Unfortunately, T$_C$ slopes for comparison sample stars with [Fe/H]$>$0.2 are not sufficiently reliable to be included in our plots. We cannot thus discuss whether this characteristic is proper to planet-host targets. More work is required to obtain more conclusive evidences on the possible cases of pollution.


\begin{references}
\reference Beirao, P., Santos, N.C., Israelian, G., et al.\ 2005, \aap, 438, 251 
\reference Ecuvillon, A.,Israelian, G., Santos, N.C., et al.\ 2004a,\aap,418,703
\reference Ecuvillon, A., Israelian, G., Santos,N.C., et al.\ 2004b,\aap,426,619
\reference Ecuvillon, A., Israelian, G., Santos, N.C., et al.\ 2005, \aap, in press, astro-ph/0509326
\reference Gilli, G., Israelian, G., Ecuvillon, A., et al.\ 2005, \aap, submitted
\reference Gonzalez, G. 1997, \mnras, 285, 403
\reference Sadakane, K, Ohkubo, Y., Takeda, Y., et al.\ 2002, PASJ,54,911
\reference Santos, N.C., Israelian, G., \& Mayor, M.\ 2000, A\&A,363,228
\reference Santos, N.C., Israelian, G., \& Mayor, M.\ 2001 A\&A,373,1019
\reference Santos, N.C.,Israelian,G., Mayor,M., et al.\ 2005,\aap,437,1127
\reference Smith, V.V., Cunha, K., \& Lazzaro, D.\ 2001, AJ,121,3207
\reference Takeda, Y. Sato, B., Kambe, E., Aoki, W., et al.\ 2001, PASJ,53,1211
\end{references}
\end{document}